\documentclass[authoryear,12pt,3p]{jowarticle}

\usepackage[english]{babel}
\usepackage[utf8]{inputenc}
\usepackage{amsmath,amsfonts,amssymb}
\usepackage{graphicx}
\usepackage[colorinlistoftodos]{todonotes}
\usepackage{natbib}
\usepackage{blindtext}
\usepackage{subcaption}
\usepackage{caption}
\usepackage{setspace}
\usepackage{tikz-cd}

\makeatletter
\renewcommand\section{\@startsection{section}{1}{\z@}{-3.25ex plus -1ex minus -.2ex}{1.5ex plus .2ex}{\normalsize\bf}}
\renewcommand\subsection{\@startsection{subsection}{2}{\z@}{-3.25ex plus -1ex minus -.2ex}{1.5ex plus .2ex}{\normalsize\bf}}
\renewcommand\subsubsection{\@startsection{subsubsection}{3}{\z@}{-3.25ex plus -1ex minus -.2ex}{1.5ex plus .2ex}{\normalsize\bf}}
\makeatother

\newtheorem{fact}{Fact}
\newtheorem{thm}{Theorem}

\begin{document}
\begin{frontmatter}
\title{A Puzzle About General Covariance and Gauge}
\author{Eleanor March}\ead{eleanor.march@philosophy.ox.ac.uk}
\address{Faculty of Philosophy \\ University of Oxford}
\author{James Owen Weatherall}\ead{james.owen.weatherall@uci.edu} 
\address{Department of Logic and Philosophy of Science \\ University of California, Irvine}

\date{\today}
%\date{ }

\begin{abstract}
We consider two simple criteria for when a physical theory should be said to be ``generally covariant'', and we argue that these criteria are not met by Yang-Mills theory, even on geometric formulations of that theory.  The reason, we show, is that the bundles encountered in Yang-Mills theory are not natural bundles; instead, they are gauge-natural.  We then show how these observations relate to previous arguments about the significance of solder forms in assessing disanalogies between general relativity and Yang-Mills theory.  We conclude by suggesting that general covariance is really about functoriality.
\end{abstract}
\end{frontmatter}

\doublespacing
\section{Introduction} \label{sec:intro}

Few observations would find more universal assent among relativists of the past century than that one should prefer -- or insist on -- generally covariant theories.  Of course, over the same period there has been little agreement about what ``general covariance'' means, much less whether it imposes a substantive constraint on physical theorizing \citep{Norton, Earman, Pooley}.  On one side are followers of \citet{Kretschmann}, who famously argued (contra Einstein) that general covariance is trivial because \emph{any} remotely plausible physical theory can be reformulated in generally covariant form; and on the other side have been dozens of attempts to identify some non-trivial, physically well-motivated principle, satisfied by general relativity and affiliated theories but not others, that captures working physicists' sense that general covariance imposes a difficult-to-meet constraint on theorizing, with real consequences for things like conservation principles \citep{Freidel+Teh}.

We will not attempt to adjudicate the many decades of dispute on this topic. (By our reckoning, it is now 11 and counting!)  Instead, we wish to identify one sense in which general covariance, or something very much like it, \emph{does} amount to a non-trivial constraint on theories---a constraint, we suggest, that is so non-trivial that it is not even satisfied by widely accepted current physics.  Specifically, we will argue that (classical) Yang-Mills theory is not generally covariant.\footnote{For background on this theory, its interpretation, and the formalism we adopt here for discussing it, see \citet{WeatherallFBYMGR}, which builds on earlier work by \citet{Palais} and \citet{Bleecker}.}  The technical facts on which this claim is based will not come as a surprise to the cognoscenti, but we suggest that their significance for foundational discussions has not been widely appreciated, in part because of lack of clarity about the meaning of general covariance.  As we will argue, that Yang-Mills theory fails to be genercally covariant has consequences for how we interpret physical geometry, the meaning of diffeomorphism invariance, the role of principal bundles and solder forms in Yang-Mills theory, and the meaning of ``gauge'' in contemporary physics.\footnote{One anonymous reviewer summarized our thesis as follows: general relativity is generally covariant, whereas gauge theories are gauge covariant.  In those terms, what we claim here may sound like a truism.  But our point, spelled out in more detail in section \ref{sec:unifying}, is that general covariance can be viewed as a special case of gauge covariance, but \emph{not vice versa}.  In other words, though every generally covariant theory is gauge covariant, not every gauge covariant theory is generally covariant.  This reverses a common view, on which gauge covariant theories satisfy both general covariance and an additional further condition related to gauge transformations.}

Of course, in order to defend this claim, we will need to say what we mean by ``general covariance''.  We will do that in section \ref{sec:genCov}.  In section \ref{sec:YM}, we will present our argument that Yang-Mills theory is not generally covariant (even though it admits a ``geometrical'' formulation and has other features quite similar to general covariance).  Then, in section \ref{sec:naturality}, we will introduce the formalism of \emph{natural bundles} \citep{Salvioli,Terng,Kolar+etal}, which extends and generalizes the geometric object program \citep{Schouten,Nijenhuis}.  As we explain there, the arguments of sections \ref{sec:genCov} and \ref{sec:YM} may be neatly summarized as follows: whereas the tangent structures to a manifold used in general relativity form natural bundles, the vector and principal bundles encountered in Yang-Mills theory are not natural.  

The remainder of the paper will address responses and consequences.  In section \ref{sec:gauge}, we will argue that the need to generalize and thereby restore naturality for Yang-Mills theories provides a new perspective on the role that principal bundles play in those theories; it also clarifies the role of the frame bundle in general relativity. In section \ref{sec:unifying}, we will discuss the relationship between natural bundles and gauge natural bundles, and how this sheds light on the sense in which gauge naturality restores a minimal sense of general covariance for Yang-Mills theory. Then, in section \ref{sec:solder}, we will argue that the foregoing arguments also provide a new perspective on the role of solder forms in some analyses of tangent structures \citep[c.f.][]{Anandan,Healey,WeatherallFBYMGR}.  Finally, we will conclude with some remarks about how naturality bears on older debates about the meaning and status of general covariance and preview some further ways in which the perspective offered here can help clarify issues in the foundations of general relativity and Yang-Mills theory.

\section{What is General Covariance?}\label{sec:genCov}

As alluded to above, general covariance is a famously vexed concept.  Many versions of general covariance, as a criterion that a physical theory may or may not satisfy, have been proposed, and many arguments have been made that those criteria are incoherent, trivial, or inadequate.  For present purposes, though, the key issues in the debate -- namely, whether some particular account does or does not provide a non-trivial and plausibly reasonable constraint -- may be set aside.  Our discussion, at least in the next several sections, concerns only certain core ideas connected to general covariance that are apparently shared among virtually all participants in the debate.  In other words, we will consider just minimal necessary conditions for general covariance; many authors have argued that (substantive) general covariance requires more than just these weak necessary conditions \citep{Norton,Pooley}.  Our goal is to show that in fact widely accepted theories fail to satisfy even these weak necessary conditions.

For our purposes, the conceptual core of general covariance is the requirement that the principal relationships posited by a theory must be preserved under coordinate transformations.  In other words, whether some physically meaningful assertion holds true cannot depend on a choice of coordinate system; were some relationship to hold when expressed in one coordinate system, then the same relationship, expressed in different coordinates using suitably transformed representations of the objects involved in the assertion, would still hold (and in fact, express the same fact). From a modern perspective \citep[c.f.][]{Misner+etal,Wald}, the requirement that some object has the appropriate transformation properties under coordinate transformations to enter into such relationships is often replaced by the requirements that the object admit a \emph{coordinate-independent} characterization and that the relationships involving such objects can be expressed using just these coordinate-independent objects.\footnote{\label{MTW} Our approach is especially close to \citet[p. 48]{Misner+etal}, who identify general covariance with the requirement that ``every physical quantity must be describable by a geometric object'' in the sense of \citet{Nijenhuis}.  They attribute the first clear articulation of this view to \citet{Veblen+Whitehead}. \citep[See also][pp. 302-3.]{Misner+etal}}  Coordinate transformations, then, may be re-interpreted as implementing diffeomorphisms acting on these coordindate-independent relationships, and invariance under coordinate changes can be interpreted as the requirement that physically meaningful relationships between coordinate-independent objects preserve their truth value under the action of diffeomorphisms.

Thus, we will assume that a theory is rightly described as generally covariant \emph{only if}:
\begin{enumerate}
    \item the objects involved in the principal claims and relationships of that theory are (or can be) expressed in a way that does not depend on particular choices of coordinate system; and 
    \item those objects have well-defined actions under diffeomorphisms, or changes of coordinate system.
\end{enumerate}
As we said above, we take these to be weak necessary conditions expected to hold of any generally covariant theory.  Any full account of general covariance would require more.  At very least, it would require that the principal claims and relationships asserted by a theory are preserved under the actions of diffeomorphisms on the objects concerned.\footnote{Much more can be said about what makes a ``theory'', understood as a system of differential equations, generally covariant in the sense suggested here, but we leave that discussion of ``natural theories'' to future work because it is not necessary for the present points. (\citet{Fatibene+Francaviglia} offer one approach for theories derived from a variational principle.)}  But even with this further requirement, we do not not claim to have captured all of the desiderata that have been required by various authors assessing whether general covariance is a plausible and substantive constraint on theories.

Consider an example.  General relativity is surely a canonical example of a generally covariant theory.  What does this mean?  Let $M$ be a smooth, four-dimensional manifold representing events in space and time, and let $g_{ab}$ be a smooth, Lorentz-signature metric on $M$ representing spatio-temporal relations between those events. (The pair $(M,g_{ab})$ will be called a \emph{relativistic spacetime} in what follows.) Finally, suppose there is some distribution of matter throughout space and time, whose energy and momentum properties can be represented by a smooth tensor field $T^{ab}$.  Then according to general relativity, Einstein's equation must hold, relating the metric and its associated curvature to the energy-momentum tensor $T^{ab}$:
\[
R_{ab} - \frac{1}{2}Rg_{ab} = k T_{ab},
\]
where $R_{ab}$ is the Ricci curvature tensor, $R$ is scalar curvature, $k$ is a constant related to Newton's gravitational constant and the speed of light, and indices are lowered using the spacetime metric.  

As we have just described it, general relativity clearly meets the two criteria we have set out.  The objects implicated in Einstein's equation -- the metric, curvature, and stress-energy tensor -- are all tensor fields, which we have presented in a coordinate independent way.  Moreover, diffeomorphisms act on all of these fields in a well-defined way, via the push-forward construction. Though it is not immediately relevant, we also note that the relationship expressed by Einstein's equation is preserved under that action by diffeomorphisms.  That is, if $\varphi:M\rightarrow N$ is a diffeomorphism between $M$ and some other manifold $N$, then we have
\begin{align*}
    R_{ab} - \frac{1}{2}Rg_{ab} = k T_{ab}& & \Leftrightarrow & &\varphi_*(R_{ab} - \frac{1}{2}Rg_{ab}) = \varphi_*(k T_{ab}),
\end{align*}
where $\varphi_*$ is the pushforward along $\varphi$. Thus, Einstein's equation is ``coordinate-independent'' in the required way, and the theory is generally covariant.

The example helps show how weak our conditions are, at least for theories formulated using tensor fields.  General covariance is automatic for such theories.  To show a theory is generally covariant, one need only rewrite it in the language of tensor calculus.  It was on these grounds that \citet{Kretschmann} argued that general covariance is trivial.  At very least, one might think he established that \emph{these} conditions are trivial, and so the whole of the dispute about general covariance is whether there are other, stronger criteria that should be imposed on top of these.\footnote{For just one example: \citep[p. 57-8]{Wald} takes for granted the two conditions required here, and adds further that ``the metric of space is the only quantity pertaining to space that can appear in the laws of physics''. Later he allows that quantities pertaining to space that are determined by the metric, such as a derivative operator, are also compatible with general covariance.}  Many arguments are available that aim to show more is required.  We set those aside because we will presently argue that even these conditions are not satisfied for realistic (and widely accepted) theories.

\section{Yang-Mills Theory is not Generally Covariant}\label{sec:YM}
Models of Yang-Mills theory consist of a relativistic spacetime $(M, g_{ab})$, a principal $G$-bundle $G\rightarrow P \rightarrow M$ over $M$, and a principal connection $\omega$ on $G\rightarrow P\rightarrow M$. Sections of vector bundles $P\times_GV \rightarrow M$ associated to this principal bundle represent matter fields that participate in the Yang-Mills interaction. Sections $\sigma:M\rightarrow P\times_GV$ of an associated bundle can, in turn, be associated with charge-current densities $J$ on $P$. These enter into the Yang-Mills equation
\begin{equation*}
    \star D\star\Omega=J
\end{equation*}
where $\star$ is the Hodge star operator relative to $g_{ab}$ (pulled back to $P$ along $\pi$), $D$ is the exterior covariant derivative relative to $\omega$, and $\Omega$ is the curvature two-form of $\omega$. For example, the structure group of electromagnetism is $U(1)$, so models of electromagnetism consist in a principal $U(1)$ bundle $U(1)\rightarrow EM \rightarrow M$ over $M$ and a principal connection $\omega$ on $EM$. In the simplest case, where matter is represented by a complex scalar field, the vector space $V$ is a copy of $\mathbb{C}$ equipped with a faithful representation of $U(1)$, and the associated bundle $EM\times_{U(1)}\mathbb{C} \rightarrow M$ has fibers isomorphic to $\mathbb{C}$.\footnote{We describe the associated bundle construction in more detail in section \ref{sec:gauge}.} 

On this way of presenting Yang-Mills theory, it does satisfy our condition (1). In particular, the objects involved in the principal claims and relationships of Yang-Mills theory, including connections on a principal bundle and sections of an associated bundle, can be characterized in a coordinate-independent way as per the above.

However, notice that the coordinates at issue here are importantly different from the coordinates at issue in the claim that general relativity admits a coordinate-independent characterization. This is because a full coordinate-based description of a section of an associated bundle, or of the connection or curvature on the principal bundle, would involve coordinates on the total space of the bundle in which they are valued, rather than the base space, i.e., the manifold representing space and time.  In general, coordinates on the base space of a bundle do not determine coordinates on the total space, and similarly coordinate transformations on the base space do not lift to coordinate transformations on the total space. 

This raises a problem for how to understand the action of smooth maps on the base space on these objects. In particular, given a diffeomorphism $\varphi:M\rightarrow M$ which acts on the base space of some vector bundle $B \rightarrow M$, and given two sections $\sigma, \sigma':M\rightarrow B$ of that bundle, there is in general no way to say whether or not those sections are ``related by the diffeomorphism $\varphi$". As a result, Yang-Mills theory does not satisfy our condition (2), at least on one plausible way of understanding what the relevant class of diffeomorphisms for condition (2) is. Note also that this failure to satisfy (2) does not depend on the details of the dynamics of Yang-Mills theory. Rather, it is a simple consequence of the fact that the theory is formulated using structures defined on a generic principal bundle and its associated vector bundles.

To make this concrete, consider the following example. Suppose we have two diffeomorphic manifolds $M$ and $N$ and (smooth) complex scalar fields on each, i.e.~smooth sections $\sigma_M$, $\sigma_N$ of the bundles $EM_M\times_{U(1)}\mathbb{C} \rightarrow M$ and $EM_N\times_{U(1)}\mathbb{C} \rightarrow N$ on $M$, $N$ respectively. Let $\varphi:M\rightarrow N$ be a diffeomorphism.  One might then ask: is $\sigma_N$ the image of $\sigma_M$ under the action of $\varphi$?  There is no way to answer this question.  The reason is that $\varphi$ does not act on points in the bundle space $EM_M\times_{U(1)}\mathbb{C}$. Indeed, there is no well-defined, unambiguous way of saying what it would mean for $\varphi$ to act on $EM\times_{U(1)}\mathbb{C}$.  What is needed is some canonical way of associating to each diffeomorphism $\varphi$ a unique bundle morphism $(\psi, \varphi)$, but in general, to do so would require further structure, such as a flat connection or a preferred global trivialization; and different choices of that additional structure would yield different associations of diffeomorphisms to bundle morphisms. This illustrates our claim that Yang-Mills theory, even in its ``geometric formulation", does not satisfy our condition (2).
 
It is worth emphasising the difference between real and complex scalar fields in this respect. Na\"ively, one might think that if this argument that Yang-Mills theory is not generally covariant works, it works too well, in that a version of this problem would also apply to smooth real scalar fields of the sort often encountered in general relativity. Recall that smooth real scalar fields, i.e.~maps $M\rightarrow\mathbb{R}$, can always be thought of as smooth (global) sections of the trivial smooth rank-$1$ vector bundle $\mathbb{R}\rightarrow B\overset{\pi}{\rightarrow}M$. So by parity of reasoning (so the thought goes): diffeomorphisms $\varphi:M\rightarrow N$ act only on the base space, but not on $B$, smooth real scalar fields are smooth (global) sections $\sigma:M\rightarrow B$, and therefore diffeomorphisms do not act on those fields. And so it would seem one cannot say whether two such fields are ``related by a diffeomorphism".

What has gone wrong here? The crucial point is that unlike the associated bundle $EM\times_{U(1)}\mathbb{C} \rightarrow M$ of Yang-Mills theory, there is always a canonical way of lifting diffeomorphisms $\varphi:M\rightarrow N$ on $M$ to bundle morphisms on $\mathbb{R}\rightarrow B\overset{\pi}{\rightarrow} M$. (As we will see in section \ref{sec:naturality}, this is deeply related to the fact that the bundle $\mathbb{R}\rightarrow B\overset{\pi}{\rightarrow} M$ is, in a certain intuitive sense, ``constructed" out of the base space $M$, whereas the bundle $EM\times_{U(1)}\mathbb{C} \rightarrow M$ is not.) To see this, first recall that any two complete ordered fields $\mathfrak{R}_1=( R_1, +_1, \times_1, \leq_1, 0_1, 1_1)$, $\mathfrak{R}_2=( R_2, +_2, \times_2, \leq_2, 0_2, 1_2)$ are uniquely isomorphic (by standard results in analysis). Now suppose that $\mathbb{R}\rightarrow B\overset{\pi}{\rightarrow}M$ is a trivial smooth rank-$1$ vector bundle. The foregoing implies that not only is $B$ isomorphic to $M\times\mathbb{R}$ (since it is a trivial bundle with $\mathbb{R}$ fibers), but there is in fact a \emph{unique} isomorphism $\chi:B\rightarrow M\times\mathbb{R}$ that preserves the complete ordered field structure on the fibers of $B$. This means that $B$ comes canonically equipped with a projection map $\pi_\mathbb{R}$ onto the second factor (since $M\times\mathbb{R}$ does, so we just pull it back by $\chi$). So we can define a canonical bundle morphism $(\psi, \varphi)$ by requiring that it preserve this projection onto the second factor, which gives us $\psi=\varphi\times\mathrm{id}_\mathbb{R}$. In other words:
whilst the diffeomorphism $\varphi$ is defined to act only on the base space, the way that we have constructed the bundle $\mathbb{R}\rightarrow B\overset{\pi}{\rightarrow}M$ means that diffeomorphisms $\varphi$ on the base space are always canonically associated with a unique action on the total space, and hence induce bundle morphisms.

This makes it clear why no analogous construction works for the bundle $EM\times_{U(1)}V \rightarrow M$. For one, the space $EM\times_{U(1)}\mathbb{C}$ need not be isomorphic to $M\times\mathbb{C}$; even if it is, we cannot use this fact to pin down a unique vector bundle morphism unless we first specify a choice of isomorphism $\chi: EM\times_{U(1)}\mathbb{C}\rightarrow M\times\mathbb{C}$, since the field $\mathbb{C}$ has a non-trivial automorphism group.

\section{Naturality, Geometricity, and Covariance} \label{sec:naturality}

As we have now seen, there is a sense in which general relativity meets certain necessary conditions for general covariance, but that Yang-Mills theory does not. What is the difference?  One way to understand what is going on here is to introduce the concept of a \emph{natural bundle}.  This is a way of capturing the idea of a ``geometric object'', developed by \citet{Schouten}, \citet{Nijenhuis} and others in the middle part of the century.\footnote{Recall fn. \ref{MTW}.}  The basic observation is that the bundles encountered in general relativity typically are natural, whereas the ones encountered in Yang-Mills theory are not natural.\footnote{Though we do not discuss them here, spinors are an example of a type of structure sometimes encountered in (extensions to) general relativity that are not natural, because to equip a manifold with a spinor structure involves an additional choice of convention \citep[c.f.][]{Fatibene+Francaviglia}. We are grateful to Henrique Gomes for raising this point, but we will postpone further discussion of it to future work.} 

Roughly speaking, natural bundles are ``species'' of bundles that depend (only) on the structure of their base space, in the sense that (a) given any (suitable) smooth manifold, one can uniquely define a bundle of the relevant species over that manifold; and (b) (suitable) smooth maps acting on base spaces ``lift'' to bundle morphisms between the natural bundles defined over them.  (The term ``suitable'' in each of these clauses will be clarified presently.)   What is intended here is clearest when one considers examples.  Take, for instance, tangent bundles.  Every manifold $M$ determines a smooth bundle $TM\rightarrow M$, whose fiber at each point is the tangent space at that point, and whose sections are smooth vector fields.  Moreover, well-behaved smooth maps $\varphi:M\rightarrow N$ between manifolds determine smooth maps between the tangent bundles of the manifolds, via the pushforward construction.  Thus, tangent bundles realize the two properties we have isolated.  Other examples abound.  Cotangent bundles, (tangent) frame bundles, (tangent) tensor bundles, bundles of metrics, bundles of $k-$forms for fixed $k$, and bundles of connections are all natural.  (So are the bundles in which real scalar fields are valued.)

The ideas just sketched can be made precise using the language of category theory.  Let $\mathcal{M}_n$ denote the category of smooth, $n-$dimensional manifolds, with smooth embeddings as morphisms.\footnote{Our definition of $\mathcal{M}_n$ here most closely follows \citet{Palais+Terng}. \citet{Kolar+etal} define the category of $n$-manifolds to have local diffeomorphisms as its arrows.  These are immersions but not necessarily injective.} Let $\mathcal{FB}$ denote the category whose objects are smooth fiber bundles and whose morphisms are smooth bundle morphisms.  Then a \emph{natural bundle} (over $n-$manifolds) is a functor $F:\mathcal{M}_n\rightarrow \mathcal{FB}$ such that (1) for every object $M$ of $\mathcal{M}_n$, $FM$ is a bundle whose base space is $M$; and (2) for every morphism $\varphi:M\rightarrow N$ of $\mathcal{M}_n$, $F\varphi$ is of the form $(\varphi_*,\varphi)$, where the maps $\varphi_*$ induces from fibers of $FM$ to fibers of $FN$ are diffeomorphisms.\footnote{Another way of doing this would be to include, as morphisms of $\mathcal{FB}$, only smooth bundle morphisms that act as diffeomorphisms on fibers.  This is the approach taken by \citet{Palais+Terng}.  Some treatments also impose a ``regularity'' condition, but a classic result due to \citet{Epstein+Thurston} establishes that that condition is automatic in the presence of the others.}  

The two informal conditions sketched above get realized in the requirements for functoriality: a natural bundle associates bundles with manifolds and bundle morphisms with embeddings.  (Moreover, it does so in a way that preserves composition and identity.)  The ``suitable'' provisos in the informal discussion, meanwhile, are made precise with our definitions of the categories, specifically with the choices of objects and arrows of $\mathcal{M}_n$.  Note that for some natural bundles one might wish to study in physics (especially, general relativity), we must modify the category $\mathcal{M}_n$ to require manifolds to satisfy further conditions.  For instance, while Lorentzian metrics have a well-defined behavior under pushforwards along embeddings, not every $n-$manifold admits any Lorentz-signature metric \citep{Geroch+Horowitz,ONeill}.

Note that the terminology is a bit odd.  A natural bundle, officially, is not a bundle at all; rather, it is a functor that associates bundles with each manifold.  This is why we wrote above of ``species'' of bundles.  Although it is an abuse of language, we will use ``natural bundle'' to refer to both natural bundle functors and to the objects in the image of those functors.  This is similar to using the phrase ``the tangent bundle'' to refer both to a general construction procedure and to specific bundles over specific manifolds that arise from that construction procedure.

We have already discussed several examples of natural bundles familiar from general relativity.  Indeed, one can easily check that all of the standard examples of fields that one encounters in relativistic field theories -- spacetime metrics, derivative operators, curvature tensors, stress-energy tensors, electromagnetic field strengths, real scalar fields, and so on -- can be seen as sections of natural bundles over spacetime.  We claim that this fact is deeply connected to the general covariance of theories involving these structures.  In fact, we take the natural bundle framework to provide a more precise specification of the two necessary conditions we identified in section \ref{sec:genCov}.  The objects under consideration in a generally covariant theory have to exhibit ``diffeomorphism'' covariance in the sense made precise by the fact that a natural bundle is functorial over smooth manifolds.

This framework also allows us to restate the claims of section \ref{sec:YM}.  Neither the principal bundles nor the associated vector bundles encountered in Yang-Mills theory are natural bundles.  In fact, there are several barriers to naturality.  One is that when we define the bundles used in Yang-Mills theory, it is common to specify only the fiber type, and not the global topology of the bundle.  We say, for instance, that we are considering an $SU(2)$ theory, which implies the fibers are $SU(2)$ torsors.  But given a base space and typical fiber, there are generally many principal bundles (respectively, vector bundles) available.  Which bundle one chooses will determine the space of global field configurations.  This failure of uniqueness, meanwhile, creates problems for defining the bundles using a functor, since a functor must assign at most one object in its codomain to each object in its domain.  

Now, admittedly, the bare technical problem can be overcome, for instance by specifying that one is considering only trivial bundles. (Though to be clear, restricting to trivial bundles does not restore naturality; it simply removes one barrier.)  This can always be done, but it comes at an interpretational cost, because it effectively rules out certain classes of global field configuration associated with non-trivial bundles (or, more generally, ones with different global topology than those in the codomain of the functor).  If one wishes to allow that field configurations associated with different bundles over a single base space are all in some sense possible configurations of a field with Yang-Mills charge, then one cannot take those fields to be sections of a (single) natural bundle over that base space.\footnote{There are deep and under-explored issues, here, about global topology of principal bundles and the physical possibility of certain global configurations of matter. We are setting those aside.  We are not aware of a physical situation in which the representational freedom afforded by using non-trivial principal bundles is needed.  Our point is only that a generic base space and typical fiber do not even uniquely determine a principal bundle, much less a way for smooth maps on the base space to lift to the total space.}

The other barriers to naturality are arguably deeper.  The bundles encountered in Yang-Mills theory are not generally constructed from (just) the structure of the base space.  This, in turn, means that there is no generally applicable and uniform -- i.e., no natural -- way to lift diffeomorphisms to act on the fibers of these bundles.  Preserving the structure of the base space is not enough to preserve the structure of the bundle.  Some further choice is needed to identify fibers that were otherwise associated with different points, and in general, that choice can be made in many different ways.  Of course, we have already made this argument in concrete detail for the case of complex scalar fields.  Now, though, we see that the problem manifests as a failure of functoriality.

\section{Gauge Naturality} \label{sec:gauge}

Thus far, we have argued that there is an important sense in which Yang-Mills theories, by virtue of being formulated on principal bundles and associated vector bundles on which base space diffeomorphisms do not act, are not generally covariant.  This situation presents a puzzle.  Surely we cannot give up on Yang-Mills theory simply because it fails to meet some abstract principle of physical theorizing, given its enormous empirical and theoretical success.  On the other hand, giving up on general covariance is also a hard bullet to bite, especially given how fundamental the two necessary conditions we identified in section \ref{sec:genCov} appear to be.  In particular, as discussed in section \ref{sec:YM}, the failure of naturality has far-reaching consequences for assessing the physical equivalence of field configurations on diffeomorphic manifolds.  This raises other questions, such as how to assess the well-posedness of partial differential equations set on such bundles.\footnote{Here is what we have in mind.  As we know from Einstein's equation, subtle issues regarding physical equivalence of solutions can arise when trying to determine whether a system of equations has unique solutions for some initial data.  Without clear criteria for equivalence of solutions, it is hard to see how to get started in analyzing uniqueness properties of those solutions.}  

For these reasons, we do not propose dropping general covariance, so much as reconsidering precisely what it demands.  As we have seen, the two necessary conditions for general covariance that we presented in section \ref{sec:genCov} can be restated as the requirement that certain structures should be natural, in the sense of being functorial.  And as we have shown, generically vector bundles are \emph{not} natural in this sense, at least over their base spaces.  But it turns out that these vector bundles can be reconstrued as natural, by adopting a different perspective on what sorts of maps should be required to lift to act on them.  This idea can be made precise using the formalism of \emph{gauge natural bundles} \citep[Ch. 12]{Kolar+etal}.\footnote{See also \citet{Fatibene+Francaviglia} for a more accessible discussion of these ideas.}  The key move is to change the category that acts as the domain of the natural bundle functor, so that the objects in that category are not the base space of the bundle under consideration but rather principal bundles, for some fixed structure group $G$, over that base space. 

We proceed similarly to as before.  We define a category $\mathcal{PB}_n(G)$ whose objects are principal $G-$bundles over $n$ dimensional manifolds and whose arrows are principal bundle morphisms whose action on the base space is a smooth embedding.  (Thus we have a full functor $B:\mathcal{PB}_n(G)\rightarrow \mathcal{M}_n$, taking each object in $\mathcal{PB}_n(G)$ to its base space, and taking arrows to their underlying smooth embedding.)  Then a \emph{gauge natural bundle} is a functor $F:\mathcal{PB}_n(G)\rightarrow \mathcal{FB}$ satisfying the following conditions: (1) the action of $F$ on objects preserves their base space, i.e., it takes principal bundles over a manifold $M$ to fiber bundles over $M$; (2) the action of $F$ on arrows preserves their action on the base space; and (3) for every object $\pi:P\rightarrow M$ of $\mathcal{PB}_n(G)$ and open set $U\subseteq M$, the inclusion arrow $(i,1_M)$, which takes the subbundle $\pi^{-1}[U]\rightarrow U$ into $P\rightarrow M$, is mapped to the inclusion arrow taking $q^{-1}[U]\rightarrow U$ into $F(\pi:P\rightarrow M)$, where $q$ is the projection map associated with $F(\pi:P\rightarrow M)$.

This definition is abstract.  The key examples of gauge natural bundles for present purposes -- that is, for the purposes of interpreting matter theories in Yang-Mills theory -- are the vector bundles associated to a principal bundle.\footnote{In fact, the construction we presently discuss is generic at ``0th order''; we discuss higher order associated bundles in the next section.}  One can construct these bundles systematically by fixing a vector space $V$ and a representation $\rho:G\rightarrow GL(V)$ of $G$ on $V$. (More generally, one can consider any fixed manifold $S$ with a left $G$ action.)  One then considers, for each principal bundle $G\rightarrow P\rightarrow M$ in $\mathcal{PB}_n(G)$, the product manifold $P\times V\rightarrow P$.  This structure can be viewed as a trivial bundle with typical fiber $V$ over $P$, though for present purposes we will keep the full product structure, so that in fact we are considering a trivial bundle with fixed global trivialization.  

The representation of $G$ on $V$, along with the right action of $G$ on $P$, determines a right action of $G$ on $P\times V$, by $(x,v)\cdot g=(x\cdot g, \rho(g^{-1})\cdot v)$ for each $g\in G$.  Let $E=(P\times V)/G$ be the smooth manifold that results by quotienting by this action, so that points of $E$ are equivalence classes $[x,v]$ of points related by the action.  Since the action of $G$ on $P\times G$ is fiber preserving over $G$, the projection $\pi:P\rightarrow G$ determines a projection $\pi_f:E\rightarrow M$.  Finally, for any smooth $G$ principal bundle morphism $f:(P\rightarrow M)\rightarrow (P'\rightarrow M')$, the map $(f,1_V):P\times V\rightarrow P'\times V$ determines a smooth bundle morphism between the quotients $(P\times V)/G\rightarrow M$ and $(P'\times V)/G\rightarrow M'$.\footnote{These quotients are the same bundles we encountered above, for which we previously used the notation $P\times_G V\rightarrow M$. Here we are emphasizing the construction procedure, and so the fact that these are quotients by a $G$ action is especially salient.}  These two constructions together can be shown to define a functor from $\mathcal{PB}(G)$ to $\mathcal{FB}$ that satisfies the conditions set forth above.  

Gauge natural bundles are like natural bundles in the sense that they associate bundles with manifolds uniformly across different manifolds, in a way that is compatible with the manifold structure (as reflected by the functoriality of the construction); and because they give rise to a notion of ``pushforward'' along maps in $\mathcal{PB}_n(G)$.  Now, though, both the assignment of bundles and the pushforward depends on more than just the base space and maps acting on base spaces; they also depend on a principal bundle over the base space and arrows between principal bundles. Why should the principal bundle structure help here?  As emphasized by various authors \citep[e.g.][]{Kolar+etal,WeatherallFBYMGR,Gomes}, a principal bundle associated with a vector bundle can be thought of as a \emph{bundle of frames}, or basis fields, for that vector bundle.  (We make this idea precise and elaborate on it below.)  Specifying information about those frames and how they transform is the missing piece in resolving the issues raised in the previous section. 

Consider, for instance, how the issue of uniqueness raised above, concerning the global topology of a non-natural bundle, is addressed here.  It remains the case that one can define many vector bundles, with different global topologies, with a given typical fiber over a generic manifold.  But that freedom corresponds exactly to the freedom to define different principal $G$ bundles with different group representations acting on those fibers over the same manifold.  Thus, by the construction above, one gets a \emph{different} vector bundle over $M$ for each principal $U(1)$ bundle over $M$. Similarly, the problem of how to define the action of diffeomorphisms on non-natural bundles is resolved by also specifying how elements of arbitrary bases at each point transform.  That extra information uniquely determines how fiber elements in an associated bundle transform.

To see what this means in more concrete detail, consider again the example of the complex scalar field discussed above.  Since the complex numbers come equipped with a preferred Hermitian product, they can be thought of as carrying a representation of $U(1)$ that preserves that product.\footnote{One could also proceed by considering principal $GL(1,\mathbb{C})$ bundles.}  (This is the ``fundamental'' representation of $U(1)$.)  To think of complex scalar fields as a natural bundle over manifolds of dimension $n$, then, one can begin with the category of $U(1)$ bundles over $n$ dimensional manifolds, $\mathcal{PB}_n(U(1))$, and then define a functor via the construction above for associated bundles, yielding, for each object $U(1)\rightarrow P\rightarrow M$ of $\mathcal{PB}_n(U(1))$, a one dimensional complex vector bundle over $M$.  Arrows in $\mathcal{PB}_n(U(1))$ specify not just an action on the base spaces of each bundle, but also specify how to identify complex phases between fibers at domain and codomain points, by specifying how bases transform.

Before proceeding, we note a connection between the remarks here and an argument from \citet{WeatherallFBYMGR}, to the effect that the principal bundles in Yang-Mills theory are ``auxiliary'' or ``secondary'' structure.  The idea is that it is the (associated) vector bundles that represent the possible states of matter, and it is the connections on those bundles that determine the physically relevant standards of constancy for those matter fields.  The principal bundles, meanwhile, serve only to coordinate the Yang-Mills connections across different types of matter that participate in the same Yang-Mills force.  They do not represent matter or its possible states directly.  Weatherall analogizes this situation to the sense in which a coach plays an auxiliary coordinating role vis a vis players on the field.  \citet{Gomes} goes even further, arguing that principal bundles are ``epiphenomenal'' in Yang-Mills theory because in fact, all matter properties can be valued in a single vector bundle, with fibers $\mathbb{C}^3\times\mathbb{C}^2\times\mathbb{C}$, and so principal bundles are not needed even to coordinate between distinct vector bundles.

These analyses bear revisiting in light of the role of principal bundles in defining gauge natural bundles.  While it is true that principal bundles do not represent possible states of matter directly, the role they do play in Yang-Mills theory is nonetheless robust and important.  What we have seen in this section is that the vector bundles in which matter takes its properties are, in the sense described above, determined by the principal bundles to which they are associated---in a way analogous to how natural bundles are determined by their base space. We also now see that what it means for matter valued in two different vector bundles to participate in the same Yang-Mills force on a given spacetime is, at least in part, for them to be images of the same principal bundle over that spacetime under two different gauge natural bundle functors.\footnote{In part, because they also need to carry the ``same'' connection.} Likewise, the bundles associated with a given type of matter across different base spaces depends not just on the base space, but also on a choice of principal bundle over that space.  

Most importantly, it is principal bundle morphisms that play the role of smooth maps on the base space in considerations of general covariance for associated vector bundles.  This means that the coordinating role of the principal bundle is not just to determine what it means for the same connection to act on different vector bundles, but also what it means to act on sections of different bundles with a single coordinate transformation or smooth (bundle) map. Even if one follows Gomes and discards principal bundles in favor of the vector bundles (and tensor bundles constructed from them) in which matter properties are valued, those vector bundles are not natural bundles.  Diffeomorphisms do not lift to them.  Further data is needed to specify how their sections behave under the action of smooth maps.  And, crucially, though that data could in principle be specified without mentioning principal bundles, it would uniquely determine (and be uniquely determined by) a gauge natural bundle functor.  We thus conclude that while principal bundles may be viewed as auxiliary or secondary in one important sense, there are other senses in which they are of central importance.

\section{A Unifying Perspective}\label{sec:unifying}

We have now seen that we can recover a sense in which Yang-Mills theory is natural, even though it is not generally covariant in the sense with which we began; and we have seen some consequences for foundational discussions about the formal structure of Yang-Mills theory.  But this discussion raises a further question.  How should we understand the relationship between natural bundles and gauge natural bundles?  There is a clear formal analogy, insofar as both are functors from some category of geometrical structures to fiber bundles associated with them.  But in fact, more can be said.  As we will presently argue, natural bundles can be seen as a special case of gauge natural bundles---or rather, they are gauge natural bundles that can be precomposed with functors from the category of n-manifolds to some category of principal $G$-bundles.  Put another way, every natural bundle functor factors through a gauge natural bundle functor.
To see this point, we will return to our motivating example of a natural bundle: the tangent bundle over $n-$manifolds, $T:\mathcal{M}_n\rightarrow \mathcal{FB}$.  As has often been observed, the tangent bundle over a given manifold can also be thought of differently, as an associated bundle to a certain principal bundle.  The relevant principal bundle in that case is the (tangent) frame bundle $GL(n,\mathbb{R})\rightarrow LM \rightarrow M$, where $GL(n,\mathbb{R})$ is the (real) general linear group in $n$ dimensions.  Then, if $V$ is an $n$-dimensional vector space equipped with a representation of $GL(n,\mathbb{R})$, the associated bundle $V\rightarrow LM\times_{GL}V \rightarrow M$ is (canonically) isomorphic to the tangent bundle $TM\rightarrow M$, and sections of $V\rightarrow LM\times_{GL}V\rightarrow M$ correspond to vector fields on $M$.  

Put in another way, these remarks show that we can think of the tangent bundle as a gauge-natural bundle $\tilde{T}:\mathcal{PB}_n(GL(n,\mathbb{R}))\rightarrow \mathcal{FB}$, precomposed with another functor $L$.  Similarly, the cotangent bundle and bundles of rank-$(r,s)$ tensor fields on $M$ can all be understood as associated to the frame bundle---and thus, as gauge natural bundles.  We can summarize the situation just presented in Fig. \ref{fig:diagram}.
\begin{figure}\centering
\begin{tikzcd}
\mathcal{M}_n \arrow[rd,"L"]\arrow[rr,"T"] & & \mathcal{FB}\\
& \mathcal{PB}_n(GL(n,\mathbb{R})) \arrow[ur, "\tilde{T}"]
\end{tikzcd}
\caption{ The tangent bundle, understood as a natural bundle functor, factors through a gauge natural bundle by precomposition with the functor $L$.\label{fig:diagram}}
\end{figure}
Note that the frame bundle functor $L$, here, can itself be seen as a natural bundle functor, once we recognize that $\mathcal{PB}_n(GL(n,\mathbb{R}))$ is a faithful subcategory of $\mathcal{FB}$.

%As in Yang-Mills theory, matter fields are represented by sections of an associated bundle; 
How can we know that \emph{every} natural bundle arises in this way?  It is a consequence of a classic result known as the \emph{finite order theorem} \citep{Palais+Terng}.  To present this result, we need to introduce another piece of machinery.\footnote{Many natural bundles of interest, such as bundles of rank-$(p,q)$ tensor fields, are associated to $LM$. But other important examples of natural bundles, such as the bundle of affine connections, or more generally the bundle of $r$-jets of sections of any first-order natural bundle, are higher-order. (The bundle of affine connections, for example, is second-order.) Since these bundles become important for thinking about differential equations on a manifold, we think that this justifies the additional generality of introducing higher order frame bundles here.} Let $M$ be a smooth (real) manifold of dimension $n$. An $r$-frame at $p\in M$ is an invertible $r$-jet $j^r_0f$, for some $f:\mathbb{R}^n\rightarrow M$ such that $f(0)=p$, where $0$ denotes the zero element of $\mathbb{R}^n$.\footnote{Recall that if $M$ and $N$ are smooth manifolds, and $f:U\rightarrow N$, $g:V\rightarrow N$ are smooth maps defined on open neighbourhoods $U$, $V$ of some $p\in M$, then $f$ and $g$ are said to be $r$-equivalent at $p$ iff they agree on all their partial derivatives up to order $r$ at $p$ (in any local coordinate charts containing $p$, $f(p)$, $g(p)$). An $r$-jet at $p$ is an equivalence class $[f]_p$ of smooth maps which are $r$-equivalent at $p$, and the $r$-jet at $p$ containing $f$ is denoted $j^r_pf$. For any smooth maps $f: U\subset M\rightarrow N$ and $g:V\subset N\rightarrow P$, if $f(p)=q$, then $j^r_qg\circ j^r_pf:=j^r_p(g\circ f)$. An $r$-jet $j^r_pf$, $f:U\subset M\rightarrow N$ is invertible iff there exists an $r$-jet $j^r_qg$, for some $g:V\subset N\rightarrow M$ satisfying $q=f(p)$, such that $j^r_qg\circ j^r_pf=j^r_p\mathrm{id}_M$ and $j^r_pf\circ j^r_qg=j^r_q\mathrm{id}_N$. To simplify notation in what follows, we will assume that if $j^r_pf$ is an $r$-jet, then $f$ is always a local smooth map, i.e.,~$f:M\rightarrow N$ denotes $f:U\subset M\rightarrow N$ for some (unspecified) open neighbourhood $U$ of $p$.  We note that our definitions of $r$ jets and $r$ frames, here, makes use of $\mathbb{R}^n$, as opposed to a generic vector space $V$, which might be viewed as more general and/or geometric (and more consistent with the style elsewhere in the paper). But for present purposes, nothing turns on this choice, and we adopt it for simplicity and consistency with the literature.} The set of all $r$-frames on $M$ $L^rM$ is a principal fiber bundle (the $r$-frame bundle) over $M$ with structure group $GL^r(n, \mathbb{R})$, where $GL^r(n, \mathbb{R})$ is the group of invertible $r$-jets $j^r_0f$, $f:\mathbb{R}^n\rightarrow\mathbb{R}^n$ with group multiplication as composition of $r$-jets, and comes with an obvious family of projection maps $\pi^r_s$, $s\leq r$, $\pi^r_s(j^r_0f)=j^s_0f$.\footnote{Our terminology follows e.g.~\citet{Kolar+etal}, though note that $L^rM$ is also sometimes called the bundle of holonomic $r$-frames (to distinguish it from, e.g.~the bundle of (non-holonomic) $r$-frames obtained by $r-1$ times recursively taking the bundle of first jets of sections of $LM$).} It is straightforward to check that $GL^1(n, \mathbb{R})=GL(n, \mathbb{R})$ and $L^1M$ is canonically isomorphic to $LM$.
Intuitively, the bundle $L^rM$ can be thought of as follows: whilst the fiber over $\pi(p)$ in the frame bundle $LM$ consists of all $1$-equivalence classes of smooth local homeomorphisms between $\mathbb{R}^n$ and $M$ at $p$, i.e.~linear isomorphisms between $\mathbb{R}^n$ and the space of $1$-equivalence classes of curves at $p$, the fiber over $\pi(p)$ in $L^rM$ consists of all $r$-equivalent smooth local homeomorphisms between $\mathbb{R}^n$ and $M$ at $p$.

With this background, we can now state the finite order theorem.
\begin{thm}
\label{thm:finiteorder}
Let $M$ be a smooth manifold of dimension $n$ and let $S$ be a smooth manifold which carries a (left) $GL^r(n, \mathbb{R})$ action. Then the associated bundle $L^rM\times_{GL^r}S\rightarrow M$ is a natural bundle, and conversely, any natural bundle over $M$ can be constructed in this way, for some $r\geq 0$.\footnote{Note that the 0th order case makes sense, but it trivializes.  The 0th order natural bundles are manifolds themselves, with 0 dimensional fibers; and the bundles of real scalar fields.}
\end{thm}
What this theorem establishes is that the facts we noted above about the tangent bundle are generic.  \emph{Every} natural bundle factors through some gauge natural bundle or other, determined by the typical fibers of the natural bundles and representations of $GL^r(n,\mathbb{R})$ on those fibers, for some $r$.  

These remarks clarify the sense in which gauge naturality restores a minimal sense of general covariance for Yang-Mills theory. As we argued in sections \ref{sec:YM} and \ref{sec:naturality}, general covariance, in the guise of naturality, is deeply related to the fact that natural bundles are, in a certain intuitive sense, `constructed from' (just) the structure of the base space. Our discussion of the finite-order theorem for natural bundles provides one way of making this notion of `being constructed out of' precise: any natural bundle is associated to some (possibly higher-order) frame bundle, and so is constructed out of the base space insofar as $r$-frame bundles are constructed out of their base space.  From this perspective, a theory (of vector fields) is ``(minimally) generally covariant'' only if its objects have a well-defined behavior under smooth maps on the base space and smoothly varying frame changes (including, possibly, higher-order frames).  What makes natural bundles distinctive, from this perspective, is that the smooth maps on the base space \emph{determine} the basis changes. This is because natural bundles are gauge natural bundles associated to principle bundles that are determined, functorially, by the base space.  For other gauge natural bundles, additional information regarding how principal bundle morphisms act on fibers is needed.

\section{Solder Forms} \label{sec:solder}

%But we have just seen that gauge natural bundles are constructed out of principal bundles in exactly the same way, i.e., any gauge natural bundle is associated to some (possibly higher-order) principal prolongation of the relevant principal bundle $P$, and so is constructed out of $P$ insofar as the $r$th-order principal prolongations of $P$ are. This illuminates the sense in which the formalism of gauge natural bundles restores a minimal sense of general covariance for Yang-Mills theory: whilst the bundles one encounters in Yang-Mills theory are not constructed out of their base spaces in the right way for sections of these bundles to have well-defined actions under diffeomorphisms, they \textit{are} constructed out of principal bundles in the right way for sections of these bundles to have well-defined actions under principal bundle morphisms. 

We now turn to an application of the ideas presented thus far.  Several authors have discussed an apparent disanalogy between Yang-Mills theory and general relativity \citep{Anandan, Healey, WeatherallFBYMGR}.  The starting point for seeing this disanalogy is effectively a restatement of the key claim of the previous section, which was that natural bundles can be seen as a special case of gauge natural bundles.  Applied to the physical theories we are considering here, this has the consequences that, much like Yang-Mills theory, general relativity can be understood as a theory of connections on principal bundles \citep{Trautman,WeatherallFBYMGR}.

But of course, as we have also seen, not all gauge natural bundles are natural bundles.  And this leads to the disanalogy between general relativity and Yang-Mills theory.  Recall, again, the diagram in figure \ref{fig:diagram}, where we see that the tangent bundle arises as an associated bundle $V\rightarrow LM\times_{GL}V\xrightarrow{\pi} M$ over the frame bundle.  We have presented this as a commuting diagram.  But in fact, to get that commuting diagram, a certain convention is adopted, regarding how to identify sections of $LM\rightarrow M$ with bases for tangent vectors.  More generally, there are many gauge natural bundle functors $\tilde{T}$ that one might consider, and for each of them there is a natural isomorphism $\theta:T\rightarrow \tilde{T}\circ L$.  Each of these natural isomorphisms realizes a sense in which the tangent bundle, over any manifold, can be seen as isomorphic to  vector bundle associated to the frame bundle over that manifold. These different possible isomorphisms are not often discussed, because the standard constructions of the tangent bundle, frame bundle, and associated bundles together determine a canonical one, relative to certain choices made in those constructions.  This canonical isomorphism equips the frame bundle with a solder form $\theta$, which is a linear isomorphism, at each $u\in LM$, between $T_{\pi(u)}M$ and $V$, that is equivariant with respect to the $GL(n,\mathbb{R})$ action on $LM$.   

As \citet{WeatherallFBYMGR} notes, this construction is very general, in the sense that \emph{any} frame bundle, including ones constructed from vector bundles that are not tangent to a manifold, comes equipped with a solder form that fixes an isomorphism between the vector bundle and associated vector bundles (of the same dimension) to its frame bundle. In more detail, let $V\rightarrow B \rightarrow M$ be any vector bundle over $M$. We can construct the frame bundle $GL(V)\rightarrow LB\xrightarrow{\wp} M$ for $B$. Since the associated bundle $LB\times_{GL}V\rightarrow M$ is isomorphic to $B$, we can equip $LB$ with an equivariant one-form that defines, at each $u\in LB$, a linear isomorphism between the fiber of $B$ at $\wp(u)$ and the fiber of $LB\times_{GL}V$ at $\wp(u)$. The construction procedures fix a preferred isomorphism, and thus a preferred solder form, in just the same way as for the tangent bundle. 

But not all principal bundles carry a (canonical) solder form.  They do so only insofar as they are viewed as (subbundles of) the frame bundle for some particular vector bundle.  And the principal bundles in Yang-Mills theory are not always thought of as frame bundles. For some authors, this fact reveals a sense in which general relativity is disanalogous to Yang-Mills theory. For example, for \citet{Anandan} and \citet{Healey}, the lack of a solder form on principal bundles shows that Yang-Mills theory does not admit the same kind of geometrical interpretation as general relativity.  Anandan even argues that the existence of the solder form ``breaks'' gauge invariance for the gravitational field.  Healey, meanwhile, takes the solder form in general relativity to partly motivate his endorsement of a holonomy interpretation of Yang-Mills theory but not general relativity. \citet{WeatherallFBYMGR}, on the other hand, drawing on the above construction, emphasizes that every principal bundle in Yang-Mills theory \emph{can} be thought of as a subbundle of a frame bundle, and in that sense does carry a solder form.  For Weatherall, the important difference is that $LM$ is soldered specifically to the tangent bundle, rather than that there is a solder form at all; and he argues that this difference does not support the conclusions drawn by Healey and Anandan.

Our discussion of naturality and gauge naturality in the previous sections provides a new perspective on this debate. In particular, we will show that once we have the formalism of natural and gauge natural bundles on the table, we can see that (i) solder forms are in some sense a generic feature of natural bundles, but that (ii) they are also, in a precisely analogous sense, a generic feature of gauge natural bundles.  And finally, (iii) the difference between solder forms in the contexts of natural bundles and gauge natural bundles has to do with the spaces that they are soldered to, rather than the fact that there exists a solder form at all. This has two conceptual payoffs. First, it clarifies the sense in which the solder form does point to a real disanalogy between general relativity and Yang-Mills theory---namely, that the bundles one encounters in general relativity are natural rather than gauge natural. Second, it provides a deeper explication of Weatherall's point that the important disanalogy between general relativity and Yang-Mills theory is not that there exists a solder form, but that $LM$ is soldered specifically to the tangent bundle. 

To establish (i)-(iii), we first need to introduce some additional machinery, similar to the notion of higher-order frame bundles discussed above.  Let $M$ be a smooth manifold of dimension $n$, and let $G\rightarrow P \overset{\pi}{\rightarrow}M$ be a principal bundle. An $r$-frame at $p\in P$ is an invertible $r$-jet $j^r_{(0,e)}f$, $f:\mathbb{R}^n\times G\rightarrow P$, $f(0,e)=p$, where $0$ again denotes the zero of $\mathbb{R}^n$ and $e$ is the identity element of $G$. The set of all $r$-frames on $P$, $W^rP$, is a principal bundle (the $r$th-order principal prolongation of $P$) over $M$ with structure group $W^r(n, \mathbb{R}, G)$, where $W^r(n, \mathbb{R}, G)$ is the group of invertible $r$-jets $j^r_{(0,e)}f$, $f:\mathbb{R}^n\times G\rightarrow\mathbb{R}^n\times G$ with group multiplication as composition of $r$-jets, and comes with an obvious family of projection maps $\pi^r_s$, $s\leq r$, $\pi^r_s(j^r_{(0,e)}f)=j^s_{(0,e)}f$.\footnote{Note that $W^r(n, \mathbb{R}, G)$ can also be defined as the semidirect product $W^r(n, \mathbb{R}, G)=GL^r(n,\mathbb{R})\rtimes J^r_nG$, where $J^r_nG$ is the (Lie) group of $r$-jets $j^r_0f$, $f:\mathbb{R}^n\rightarrow G$ with group multiplication defined via $j^r_0f\circ j^r_0g:=j^r_0(f\cdot g)$, where $\cdot$ here denotes group composition in $G$ \citep[see][Ch. XII]{Kolar+etal}.} 

We will now turn to claims (i)-(iii). To establish point (i), we now make use of Theorem \ref{thm:finiteorder} along with the following fact.
\begin{fact}\label{fact:canonicalformsframe}
    Let $M$ be a smooth manifold of dimension $n$ and $V$ an $n$-dimensional vector space. Then every $L^rM$ carries a canonical equivariant one-form $\theta^r$, which assigns, to each $p\in L^rM$, a linear isomorphism between $T_{\pi^r_{r-1}(p)}L^{r-1}M$ and $V\oplus\mathfrak{gl}^{r-1}(n,\mathbb{R})$.
\end{fact}
Fact \ref{fact:canonicalformsframe} can be seen as follows. First, notice that, like the frame bundles discussed in the previous section, $LM\rightarrow M$, for any principal bundle $P\rightarrow M$, $W^1P$ comes equipped with a canonical form $\theta_G$. This is because, if $V$ is an $n$-dimensional vector space, then the associated bundle $W^1P\times_{W^rG}(V\oplus\mathfrak{g})$, where $\mathfrak{g}$ is the Lie algebra of $G$, is isomorphic to the bundle $TP/G$ \citep[p.~155]{Kolar+etal}. Again, there are many such isomorphisms, but the construction of the associated bundle and quotient tangent bundle determine a canonical one, relative to certain choices made in those constructions. This canonical isomorphism equips $W^1P$ with a canonical form $\theta_G$, which is a linear isomorphism, at each $u\in W^1P$, between $T_{\pi(u)}P$ and $V\oplus\mathfrak{g}$. Now observe that for any $r\geq 1$, $L^rM\rightarrow M$ is a principal bundle, and so we can apply the principal prolongation construction just described. It follows that $W^1(L^{r-1}M)$ carries the canonical form $\theta_{GL^{r-1}}$.  Since every local smooth map $f:\mathbb{R}^n\rightarrow M$ lifts to a local smooth map $j^rf:\mathbb{R}^n\times GL^r(n, \mathbb{R})\rightarrow L^rM$ in the neighbourhood of the identity element $e^r$ of $GL^r(n, \mathbb{R})$,\footnote{Defined via the condition $j^r(j^{r}_pg)=j^{r}_p(f\circ g)$ for all $g:\mathbb{R}^n\rightarrow \mathbb{R}^n$.} we can pull $\theta_{GL^{r-1}}$ back to $L^rM$ via the map $L^rM\rightarrow W^1(L^{r-1}M)$, $j^r_0f\rightarrow j^1_{(0,e^{r-1})}j^{r-1}f$ to obtain $\theta^r$.\footnote{Note that this makes sense, since the element $j^1_{(0,e^{r-1})}j^{r-1}f$ depends only on $j^r_0f$.} Putting Theorem \ref{thm:finiteorder} and Fact \ref{fact:canonicalformsframe} together, we see that solder forms are a generic feature of natural bundles, in the sense that any natural bundle is associated to a bundle which carries a solder form.\footnote{To be slightly more precise, we have established this claim for natural bundles of order $\geq 0$.  The 0th order case consists of natural bundles that are not gauge natural over the frame bundle, because they are just bundles of scalar fields.}

We now move on to establish point (ii).  The key point here is that there is a version of Theorem \ref{thm:finiteorder} for gauge natural bundles as well, originally due to \citet{Eck} and strengthened by \citet{Kolar+etal}): 
\begin{thm}\label{thm:gaugefiniteorder}
    Let $G\rightarrow P\overset{\pi}{\rightarrow}M$ be a principal bundle and let $S$ be a smooth manifold with a (left) $W^r(n, \mathbb{R}, G)$ action. Then the associated bundle $W^rP\times_{W^rG}S\rightarrow M$ is a gauge natural bundle, and conversely, any gauge natural bundle over $P$ can be constructed in this way, for some $r\geq 0$.\footnote{Note, as in Theorem \ref{thm:finiteorder}, that the 0th order case makes sense.  But in this case, it does not quite trivialize, because the standard associated bundle construction yields a 0th order gauge natural bundle.  This reflects a difference in order counting between natural and gauge natural bundles.  For instance: the frame bundle and tangent bundle are both first order natural bundles, but they are 0th order gauge natural bundles.\label{order}}
\end{thm}
One also has an analogue to fact \ref{fact:canonicalformsframe} 
\begin{fact}\label{fact:canonicalformsgauge}
    Let $G\rightarrow P\overset{\pi}{\rightarrow}M$ be a principal bundle over a smooth manifold $M$ of dimension $n$ and $V$ an $n$-dimensional vector space. Then every $W^rP$, for $r\geq 1$, carries a canonical equivariant one-form $\theta_G^r$, which assigns, to each $p\in W^rP$, a linear isomorphism between $T_{\pi^r_{r-1}(p)}W^{r-1}P$ and $V\oplus\mathfrak{w}^{r-1}(n, \mathbb{R}, G)$.
\end{fact}
The argument is much like before. We have seen, by the argument above, that $W^1P$ carries a canonical form.  Then, for $r\geq 1$, since $W^1(W^{r-1}P)$ carries the canonical form $\theta_{W^{r-1}G}$, and every local smooth map $f:\mathbb{R}^n\times G\rightarrow P$ lifts to a local smooth map $j^rf:\mathbb{R}^n\times W^r(n, \mathbb{R}, G)\rightarrow W^rP$ in the neighbourhood of the identity element $e^r_G$ of $W^r(n, \mathbb{R}, G)$, we can pull $\theta_{W^{r-1}G}$ back to $W^rP$ via the map $W^rP\rightarrow W^1(W^{r-1}P)$, defined by $j^r_{(0,e)}f\mapsto j^1_{(0,e)}j^{r-1}f$ to obtain $\theta^r_G$. So again, putting Theorem \ref{thm:gaugefiniteorder} and Fact \ref{fact:canonicalformsgauge} together, solder forms are also a generic feature of gauge natural bundles, in the sense that any gauge natural bundle is associated to a bundle which carries a solder form. 

These observation do not quite complete the argument, however, and a further clarification is called for.  We have shown that gauge natural bundles of order $r\geq 1$ carry a canonical form.  But what about bundles of order $0$?\footnote{Recall the clarification about order counting in note \ref{order}.  The tangent bundle is order 1 as a natural bundle, but order 0 as a gauge natural bundle, and so the arguments we give presently apply to that as well.}  Those are the bundles we encounter most often in Yang-Mills theories, as associated vector bundles to principal bundles.  The construction we have given here does not apply to those.  Instead, those gain a solder form by the argument given by \citet{WeatherallFBYMGR}, once we identify the principal bundle to which they are associated as a subbundle of their frame bundle.  This is possible, and can be done canonically via the definitions of the structures in question, for any associated vector bundle determined by a faithful representation.  The point of the argument above is to show that once one has such a solder form for a 0th order bundle, solder forms exist for all higher order jet bundles over that vector bundle, in a way exactly analogous to how higher order natural bundles comes equipped with solder forms.

Finally, this takes us onto point (iii), and our two conceptual payoffs. As we have just seen, one can construct a variety of solder forms for both natural and gauge natural bundles. From this perspective, the fact the solder form is `more apparent' in general relativity than in Yang-Mills theory is just an artefact of the fact that many of the bundles one encounters in Yang-Mills theory are $0$th order, whereas many of the bundles one encounters in general relativity are $1$st-order. This does not mean that there is not a real disanalogy between solder forms in the contexts of natural and gauge natural bundles. As we have now seen in some detail, and as Weatherall notes, the disanalogy is that, for example, $LM$ is soldered to $TM$, whereas $W^1P$ is soldered to $TP/G$. But once we have the formalism of natural and gauge natural bundles on the table, we can see that this disanalogy is really just a manifestation of the fact that the bundles one encounters in general relativity are natural, whereas the bundles one encounters in Yang-Mills theory are gauge natural. 

\section{Conclusion: General Covariance Revisited}\label{sec:conclusion}

Our primary goal in this paper was to extract, from the long and vexed literature on general covariance, certain precise necessary conditions; and then to argue that those conditions are non-trivial because in fact they are violated by Yang-Mills theory.  We then argued that something like general covariance, properly generalized, could be restored for Yang-Mills theory, but only by moving to a more general mathematical setting; and we showed how this perspective could provide fruitful insight into other debates, concerning the status and significance of principal bundle and solder forms in understanding the relationship between general relativity and Yang-Mills theory.  

We have not attempted to give a general or complete account of general covariance.  As we have noted, to do so, we would need to say much more about what it means for a \emph{theory} to have the right sort of behavior under diffeomorphisms, whereas our focus has been on properties of the objects that appear in our theories.  We postpone that more complete discussion for future work.  Even so, we will conclude by suggesting that the (somewhat preliminary) arguments and observations here offer a valuable perspective on what general covariance and related issues, such as coordinate-independence, are really about.

General covariance is often discussed in terms of coordinate transformations and coordinate independence.  But we take the lesson of the forgoing discussion to be that more important than coordinate independence is how the objects we use in constructing physical theories depend on one another---and, in particular, how those objects depend on a spacetime manifold and other, related structures.  In fact, coordinate independence turns out not to be fully probative, because as we have seen, objects can have coordinate independent characterizations without having well-defined behavior under coordinate transformation (or, better yet, smooth maps).  This is because objects can be coordinate-independent simply because they do not have the right relationship with a manifold to be candidates to depend on (manifold) coordinates.  The complex vector bundles we discussed above demonstrate this.

We can restore a minimal sense of general covariance for Yang-Mills theory by recognizing that the background geometry of Yang-Mills theory consists of \emph{only} the structure of a principal bundle, along with a family of vector bundles, all systematically related to one another, endowed with certain further structure preserved under the action of the principal bundle's structure group: an inner product and, in some cases, an orientation.  No further structure, such as a preferred trivialization or coordinatization, is assumed.  In this way, general covariance for Yang-Mills theory is analogous to general covariance for general relativity, where the background geometry consists of \emph{only} the structure of a smooth manifold, along with various bundles that can be constructed from that.  Of course, to represent concrete situations in Yang-Mills theory, one adds to this background structure sections of certain gauge natural bundles, such as a connection and vector fields representing matter configurations.  But this is also analogous to general relativity, where one introduces a metric and various matter fields, all sections of natural bundles.

Talk of ``structures'' and ``dependence'' is evocative, but not very precise.  The key idea in the present case is explicated using the formalism of natural and gauge-natural bundles.  We suggest that it is naturality, or more precisely, still, functoriality, that captures the core of what general covariance is concerned with.  The difference between the sorts of objects and relationships we encounter in general relativity -- the generally covariant ones -- and those in Yang-Mills theory is that the former, but not the latter, involve objects that depend only on a manifold that represents space and time; whereas the latter depends on further structure on top of that.  Functoriality reflects this idea by enforcing the requirement that our constructions can be applied uniformly across manifolds, in a way that is compatible with those maps that we take to preserve manifold structure.  Coordinate- or frame-dependent constructions fail to be functorial (over manifolds) because they require further information that is not generally preserved by diffeomorphisms. This is why those constructions fail to be generally covariant.  But that is not the only way to fail to be generally covariant, and the natural bundle formalisms shows very clearly why.

\section*{Acknowledgments}
This material is based upon work supported by the National Science Foundation under Grant No. 2419967.  JOW is grateful to Henrique Gomes and David Malament for conversations on material related to this paper. EM acknowledges the support of Balliol College, Oxford, and the Faculty of Philosophy, University of Oxford.  We are grateful to Henrique Gomes for comments on a previous draft.

\bibliographystyle{elsarticle-harv}
\bibliography{covariance} 

\end{document}